\documentstyle[12pt,epsf]{article}
\def\be{\begin{equation}}
\def\ee{\end{equation}}
\def\l{\label}
\newcommand{\plot}[1]
{\begin{center} \epsfxsize=7cm
\parbox{\epsfxsize}{\epsffile{#1}}
\end{center}}
\def\tm{$^{\mathrm TM}$}
\def\kcal{\mbox{ kcal}}
\def\gas{\mbox{ (gas) }}
\def\liquid{\mbox{ (liquid) }}
\def\solid{\mbox{ (solid) }}
\def\plasma{\mbox{ (plasma) }}
\def\C{$^o\mbox{C}$}
\def\eV{\textrm{ eV}}
\def\kcal{\textrm{ kcal}}

\def\O{\mbox{\rm O}}
\def\H{\mbox{\rm H}}
\def\C{\mbox{\rm C}}

\def\be{\begin{equation}}
\def\ee{\end{equation}}
\begin{document}
\begin{titlepage}
\title{\large\bf
A STUDY OF THE ENERGY EFFICIENCY OF HADRONIC REACTORS
OF MOLECULAR TYPE}
\author{\bf A.K. Aringazin$^{1,2}$ and R.M.
Santilli$^2$}
\date{\normalsize
{$^1$Institute for Basic Research, Department of Theoretical
Physics,\\ Eurasian National University, Astana 473021
Kazakstan}\\[0.5cm]
{$^2$Institute for Basic Research, P.O. Box 1577, Palm
Harbor,}\\
{FL 34682, USA}\\
{\tt ibr@gte.net}\\[0.5cm]
{\small June 5, 2000; Revised October 10, 2001\\
Final version December 9, 2001} }
 \maketitle

\abstract{In this paper, we introduce an estimate of the
"commercial efficiency" of Santilli's hadronic reactors\tm{} of
molecular type \cite{1} (Patented and International Patents
Pending) which  convert a liquid feedstock (such as automotive
antifree and oil waste, city or farm liquid waste, crude oil,
etc.) into the clean burning magnegas\tm{} plus heat acquired by
the liquid feedstock. The conversion is done via a new process
based on a certain flow of the liquid feedstock through a
submerged electric arc between carbon-base electrodes and other
features.  The "commercial efficiency" is defined as the ratio
between the total energy output (energy in magnegas plus heat) and
the electric energy used for its production, while the "scientific
efficiency" is the usual ratio between the  total energy output
and the total energy input (the sum of the electric energy plus
the energy in the liquid feedstock as well as that in the carbon
electrodes). Needless to say, the scientific efficiency is always
smaller than one because of the conservation of energy. However, a
peculiar features of Santilli's hadronic reactors of molecular
type is that their commercial efficiency is considerably bigger
than one, namely, the reactors are capable of tapping energy from
the liquid feedstock and the carbon rods.  A primary purpose of
this paper is to show that conventional thermochemistry does
indeed predict a commercial efficiency bigger than one, although
their values is considerably smaller than the actual efficiency
measured in the reactors, thus indicating the applicability of the
covering hadronic chemistry from which the reactors have received
their name. Under the conditions that the reactions run at
temperature $T= 25^o$ C and pressure $p=1$ atm, the chemical
composition of the combustible gas is conventional, and all
thermochemical calculations processes are conventional,  we reach
an upper limit of the commercial efficiency of 3.11 for the use of
pure water as feedstock, and of 3.11 to 7.5 for a mixture of
ethyleneglicole and water with increasing relative consumption of
ethyleneglicole with respect to the consumption of carbon rods.
The study of the heat produced by the reactions leads to large
divergencies between the thermochemical predictions and
experimental data of at least a {\sl factor of three}. Such
divergencies can only be explained with deviations from quantum
chemistry in favor of the covering hadronic chemistry and. In
particular, the indicated large divergencies can only be explained
with the assumption that the produced combustible gas has the new
non-valence chemical structure of Santilli magnecules.}

\end{titlepage}

\section{Introduction}

In this paper, we introduce an estimate of the "commercial
efficiency" of Santilli's hadronic reactors\tm{} of molecular type
\cite{1} (Patented and International Patents Pending) which
convert a liquid feedstock into the clean burning magnegas\tm{}
plus heat acquired by the liquid feedstock.

The reactors operate via a new process based on a certain flow of
the liquid feedstock through a submerged electric arc between
submerged carbon-base electrodes (for which reason the reactors
are also called PlasmaArcFlow\tm{} reactors) and other features.

The "commercial efficiency" \cite{1} is defined as the ratio
between the total energy output (energy in magnegas plus heat) and
the electric energy used for its production, while the "scientific
efficiency" is the usual ratio between the  total energy output
and the total energy input (the sum of the electric energy plus
the energy in the liquid feedstock as well as that in the carbon
electrodes). Unless otherwise specified, the word "efficiency" is
referred hereon to the "commercial efficiency." The latter name
originates from the fact that liquid wastes carry an income,
rather than having a cost and, for this reason, they are not
included in commercial calculations of operating costs.

Needless to say, the scientific efficiency is always smaller than
one because of the conservation of energy. However, a peculiar
feature of Santilli's hadronic reactors of molecular type is that
their commercial efficiency is considerably bigger than one,
namely, the reactors are capable of tapping energy from the liquid
feedstock and the carbon rods.

A primary purpose of this paper is to show that conventional
thermochemistry does indeed predicts a commercial efficiency
bigger than one, although their values is considerably smaller
than the actual efficiency measured in the reactors, thus
indicating the applicability of the covering hadronic chemistry
from which the reactors have received their name.

By its elementary chemical content, magnegas is similar to the
water gas, or synthesis gas, although we should emphasize that
magnegas is produced under a DC electric arc, and reveals an
unusual chemical structure characterized by the presence of heavy
molecular mass clusters, which have not been identified by Gas
Chromatography Mass-Spectroscopy and InfraRed spectroscopy
(GC-MS/IR) tests among about 135,000 species \cite{1,2}. This
feature may be naturally attributed to the influence of the plasma
arc and related strong external magnetic field which can lead to
new couplings of CO and H$_2$ molecules and other new effects.

In Secs.  2 and 3, we consider in detail conventional chemical
reactions in a PlasmaArcFlow reactor operating with pure water or
ethyleneglicole and water mixtures as feedstock. We treat the gas
produced as a simple mixture of carbon monoxide CO and hydrogen
H$_2$ viewed as ideal gases, to simplify consideration, and
calculate its combustion heat.

We estimate the upper theoretical limit of the efficiency of the
reactor by using only chemical energy balance equations. The
efficiency is defined as a ratio between the total energy release
(including combustion heat of magnegas) to the energy input
(electricity consumed). Such efficiency is over unity due to the
fact that the sum of the combustion heat of the gas and the heat
acquired by the liquid is bigger than the electric energy needed
for their production. It is therefore evident that, for the case
of water as feedstock, the missing energy originates from the
combustion of carbon with oxygen originating from the separation
of water. This is due to the fact that the original water is
reproduced in the combustion and, therefore, cannot contribute to
the total efficiency.

Independent experimental tests of the efficiency of the
PlasmaArcFlow reactors clearly confirm such a commercial
over-unity \cite{1}, since the measured value of the over-unity is
of about 3 to 5 for antifreeze stock at atmospheric pressure with
bigger values for bigger pressures and powers. Our theoretical
result is that the upper limit of the commercial over-unity ranges
from 3.11 to 7.5, in a remarkable correspondence to the tests.

However, it should be noted that, whenever the study is specified
to the heat acquired by the liquid feedstock a discrepancy of a
factor of three originates between experimental data and the
prediction of thermochemical calculations. An additional
discrepancy also of a factor of about three exists between the
measured combustion heat of magnegas and its predicted value.

The above discrepancies are of such a magnitude to support the
hypothesis that the chemical composition of magnegas is that
Santilli's magnecules~\cite{1}.

It should also be noted that our calculations are based, as usual,
on thermochemical values at $T= 25^o$C and pressure $p=1$ atm
while the arc plasma (reaction zone) is characterized by much
higher temperatures. Therefore, our results are of preliminary
character. Also, in the present paper we do not consider issues
pertaining to mechanism of the reactions, thermodynamics,
fluidodynamics, and chemical kinetics.

In Sec.~4 we consider in detail the energy balance for the plasma
creation. In Sec.~5 we introduce the heat production coefficient
and calculate the heat production. In Sec.~6 we consider the total
heat production as a sum of the heat produced and the combustion
heat of the gas. In Sec.~7 we outline the results. Numerical data
are presented in Appendix.

\section{Water as a feedstock of PlasmaArcFlow reactors}

The main chemical reactions in PlasmaArcFlow reactors are the
gasification of carbon (graphite), evaporation of water, and the
conversion of water and carbon to hydrogen and carbon monoxide,
according to the known reactions
\be
\mbox{C(solid)} \to \mbox{C(gas)} - 171.7,
\ee
\be
\mbox{H$_2$O(liquid)} \to \mbox{H$_2$O(vapor)} - 10.4,
\ee
\be
\mbox{H$_2$O(vapor)}+\mbox{C(gas)} \to
\mbox{H$_2$(gas)}+\mbox{CO(gas)}+138.8,
\ee
in kcal/mol. Therefore the related the balance
reaction, i.e.,
\be\l{water}
\mbox{H$_2$O(liquid)}+\mbox{C(solid)}\to
\mbox{H$_2$(gas)}+\mbox{CO(gas)}-43.9,
\ee
is endothermic. Hereon, we use binding energies represented in
Table~1.
\begin{table}
\begin{center}
\begin{tabular}{|cc||cc|}
\hline
Diatomic molecules &&Diatomic molecules &\\
\hline
H--H              &104.2 & C=O & 255.8\\
\hline
O=O              &119.1 & N$\equiv$N &192.0\\
\hline
\hline
Manyatomic molecules &&Manyatomic molecules &\\
\hline
C--O                 &85.5    & O--H & 110.6\\
\hline
C=O in CO$_2$       &192.0   & O--O & 35\\
\hline
\end{tabular}
\caption{Binding energies, kcal/mole. $T=25^o$C.}
\l{Table1}
\end{center}
\end{table}

The energy input of 1~kW$\cdot$h = 860 kcal produces 860/43.9 =
19.6 moles = 19.6$\times$22.4 l = 439 l = 439/28.317 cf = 15.5 cf
of H$_2$ and the same amount of CO, treated here as ideal gases (1
mole = 22.4 l); conversion factors are presented in Table~2.
\begin{table}[ht]
\begin{center}
\begin{tabular}{ll}
\hline\\
1 kcal &= 3.9685 BTU\\
1 kcal &= 1.1628$\times$10$^{-3}$ kW$\cdot$h\\
1 BTU  &= 0.25198 kcal\\
1 BTU  &= 2.930$\times$10$^{-4}$ kW$\cdot$h\\
1 kW$\cdot$h &= 3413.0 BTU\\
1 kW$\cdot$h &= 859.99 kcal\\
1 m$^3$      &= 35.314 cf\\
1 cf         &= 28.317 liters\\
\hline
\end{tabular}
\caption{Conversion factors.}
\l{Table2}
\end{center}
\end{table}

The combustion of the products is exothermic,
\be\l{CO}
\mbox{CO} + \frac{1}{2} \mbox{O$_2$} \to
\mbox{CO$_2$(gas)} + 68.7,
\ee
\be\l{H2}
\mbox{H$_2$} + \frac{1}{2}\mbox{O$_2$} \to
\mbox{H$_2$O(vapor)} + 57.5,
\quad \mbox{H$_2$} + \frac{1}{2}\mbox{O$_2$} \to
\mbox{H$_2$O(liquid)}
+ 67.9,
\ee
\be
\mbox{CO} + \mbox{H$_2$} + \mbox{O$_2$} \to
\mbox{CO$_2$} + \mbox{H$_2$O(vapor)} + 126.2,
\ee
\be
\mbox{CO} + \mbox{H$_2$} + \mbox{O$_2$} \to
\mbox{CO$_2$} + \mbox{H$_2$O(liquid)} + 136.6,
\ee
i.e., the 50\%-50\% mixture of (CO+H$_2$) ideal gas has 68.3
kcal/mol = 271.2 BTU/mol = 342.8 BTU/cf
content\footnote{Due to experimental tests magnegas, which
consists mainly of CO and H$_2$ at approximately equal
percentages, has about 800 BTU/cf energy content \cite{1}.}. The
total combustion heat is 19.6$\times$126.2 = 2473.5 kcal = 2.88
kW$\cdot$h (for water vapor) and 19.6$\times$136.6 = 2677.4 kcal =
3.11 kW$\cdot$h (for water liquid), respectively. Therefore, the
theoretical upper limit of the efficiency is 3.11.

Clearly, only some part $k$ of the consumed electric energy
contributes directly the reaction (\ref{water}) because some of
the electric energy is consumed in the production of heat
(dissipation). Consequently, the real efficiency is $3.11k$, where
$k<1$.

\section{Antifreeze as a feedstock of PlasmaArcFlow
reactors}

We assume that antifreeze consists of ethyleneglicole and water.
The complete dissociation of ethyleneglicole (we ignore
evaporation heat and solution effects) is characterized by
\be
\mbox{HO-CH$_2$-CH$_2$-OH(gas)} \to
4\mbox{C} + 6\mbox{H} + 2\mbox{O} - 869.6,
\ee
and the subsequent association of CO and H$_2$,
\be\l{et}
\mbox{HO-CH$_2$-CH$_2$-OH} \to 2\mbox{CO}+
3\mbox{H$_2$} - 45.4,
\ee
produces 2 moles of CO and 3 moles of H$_2$.

In PlasmaArcFlow reactors we thus have a pair of chemical
reactions, (\ref{water}) and (\ref{et}), or
\be\l{etwater}
r\mbox{HO-CH$_2$-CH$_2$-OH}  + \mbox{H$_2$O} +
\mbox{C} \to
(2r+1)\mbox{CO}+ (3r+1)\mbox{H$_2$} - (45.5r+43.9),
\ee
where $r$ represents the relative consumption of ethyleneglicole
with respect to that of carbon rod.

The energy effect of reaction (\ref{etwater}) is endothermic,
-(45.4$r$ + 43.9) kcal $<0$. The energy input of 1~kW$\cdot$h =
680 kcal produces $(2r+1)680/(45.4r+43.9)$ moles of CO and
$(3r+1)680/(45.4r+43.9)$ moles of H$_2$, with the total combustion
energy,
\be
\frac{680}{45.4r+43.9}\left((2r+1)68.7 + (3r+1)67.9\right)
\mbox{kcal},
\ee
where we have used Eqs. (\ref{CO}) and (\ref{H2}) Therefore, the
upper theoretical limit of the efficiency is given by
\be\l{rless}
\frac{1}{45.4r+43.9}\left((2r+1)68.7 +
(3r+1)67.9\right).
\ee
For $r=0$, we recover the value 3.11 obtained in Sec.~2. The
efficiency increases from 3.11 to 7.51 with the increase of $r$
from 0 to infinity. Figures 1 and 2 display efficiency
(\ref{rless}) as a function of $r$. Only some part $k$ of the
electric energy consumed contributes directly the reaction
(\ref{etwater}) because, again, some electric energy is dissipated
into heat. Therefore, the real efficiency of the reactor is less
than that given by Eq. (\ref{rless}).
\begin{figure}
\plot{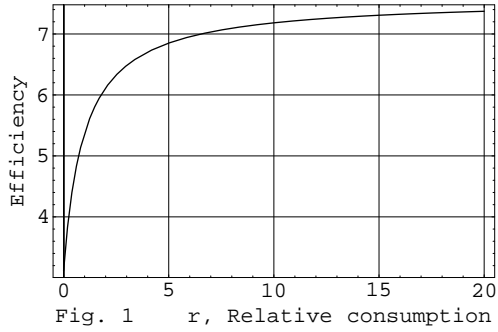} \caption{Theoretical efficiency of PlasmaArcFlow
reactors as a function of relative consumption $r$ of
ethyleneglicole with respect to carbon rods; $0\le r\le 20$.}
\label{Fig1}
\end{figure}
\begin{figure}
\plot{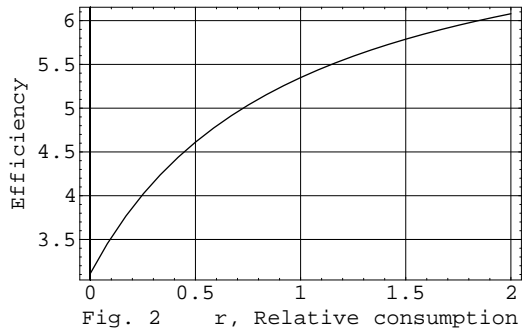} \caption{Theoretical efficiency of PlasmaArcFlow
reactors as a function of relative consumption $r$ of
ethyleneglicole with respect to carbon rods; $0\le r\le 2$.}
\label{Fig2}
\end{figure}

In general, higher consumption rates of ethyleneglicole and carbon
rods per 1 kW$\cdot$h electricity consumed imply bigger real
efficiency. This is due to higher values of $k$, which depend on
design of the reactor. In turn, the consumption rates depend on
reaction rates, volume of the reaction zone, rates of the
reactants (ethyleneglicole, water, gasified carbon rod) input,
rates of the products (CO and H$_2$) removal, stoichimetric
ratios, etc. The reaction rates depend on temperature and
pressure. The volume of the reaction zones depends on size of the
plasma arc and on the size of surrounding high-temperature
regions. The rates of the reactants inlet and products outlet
depend on the rate of the carbon rod gasification, geometry and
velocity of the liquid flow, and pressure. Here, it is important
to identify a limiting factor (e.g., the slowest rate among the
above) in order to better represent the efficiency of the reactor.

\section{Accounting for plasma creation}

We now present calculations of the energy required to convert
liquid water and solid graphite into the plasma state so as to
identify its possible contribution to the overall efficiency.

\subsection{Water contribution}

We take 1 mole of liquid water at $T=20^o \C$ and atmospheric
pressure as an initial state.

(1) The energy required to heat up one mole of water from $T=20
\C$ to $T=100 \C$ is $Q_1 = 1.4$ kcal.

(2) The energy required to evaporate one mole of water is $Q_2 =
10.4$ kcal.

(3) The energy required to heat up one mole of water vapor from
$T=100 C$ to $T=3600 C = 3300 K$ is $Q_3 = 26$ kcal.

(4) The energy required for total disintegration of one mole of
water molecules to individual atoms is $Q_4 = 221.6$ kcal,
\be
\H_2\O \to \H+\H+\O -Q_4.
\ee

(5) The energy required to ionize all the H and O atoms can be
calculated due to the following known values of the first
ionization potentials,
\be
 \H \to \H^{+} + e - 13.6 \eV, \quad
 \O \to \O^{+} + e - 13.6 \eV.
\ee
Taking into account that  1 eV = $3.83 \cdot 10^{-23}$ kcal, the
Avogadro number is $N = 6 \cdot 10^{23}$ particles per mole,  1
mole of water (i.e. $N$ molecules of water) gives $2N$ atoms of
the hydrogen and $N$ atoms of the oxygen (in total $3N$ atoms), we
have
\be
Q_5 = 3N 13.6 \eV = 3\cdot6\cdot10^{23}\cdot13.6 \eV =
244.8\cdot10^{23} \eV,
\ee
 i.e.,
\be
Q_5 = 244.8\cdot10^{23}\cdot(3.83 \cdot 10^{-23}) \kcal = 937.6
\kcal.
\ee

In total we obtain the following energy required to convert 1 mole
of liquid water into the pure plasma state,
\be
\H_2\O(\textrm{liquid}) \to \H_2\O(\textrm{plasma}) -
Q,
\ee
where
\be
Q = Q_1+Q_2+Q_3+Q_4+Q_5,
\ee
so that by inserting the above values, we finally get the
following numerical value:
\be
Q = 1.4 + 10.4 + 26 + 221.6 + 937.6 = 1197 \kcal.
\ee

\subsection{Carbon contribution}

We assume 1 mole of solid carbon (graphite) at $T=20^o C$ and
atmospheric pressure as an initial state.

(1) The energy required to heat up one mole of graphite from
$T=300 K$ to $T=3300 K$ is $E_1 = 6$ kcal;

(2) The energy required to evaporate one mole of graphite,
$\C\solid \to \C\gas - E_2$, is $E_2 = 171.7$ kcal;

(3) The energy required to ionize one mole of graphite can be
calculated due to the following known values of the first
ionization potential:
\be
\C \to \C^{+} + e - 11.3 \eV.
\ee
One mole contains $N=6\cdot10^{23}$ atoms, so the
require energy is
\be
E_3 = 11.3N \eV = 67.8 \cdot 10^{23} \eV,
\ee
or, using 1 eV = $3.83 \cdot 10^{-23}$ \kcal,
\be
E_3 = 259.7 \kcal.
\ee
In total, we obtain the following energy required to convert 1
mole of solid carbon to the pure plasma state,
\be
\C(\textrm{solid}) \to \C(\textrm{plasma}) - E,
\ee
where
\be
E = E_1+E_2+E_3,
\ee
so that by inserting the above values, we finally obtain the
following numerical value,
\be
E
= 437.4 \kcal.
\ee

\subsection{Fully ionized plasma of 2H, O and C}

In total, the energy required to convert 1 mole of liquid water
and 1 mole of solid carbon to a pure plasma state is the sum of
the above two energies,
\be
W = Q + E,
\ee
i.e.,
\be
W
= 1197 + 437.4 = 1634.4 \kcal.
\ee
This energy is required to convert 1 mole the water and 1 mole of
carbon to 4 moles of the pure plasma, as a sum of 2 moles of H, 1
mole of O, and 1 mole of C. More precisely, the plasma consists of
$2N$ positive ions H$^{+}$, $N$ positive ions O$^{+}$, $N$
positive ions C$^{+}$, and $4N$ electrons.

We can convert moles to cubic foots by assuming that the plasma is
an ideal gas. Using the facts that 1 mole of ideal gas is 22.4
liters and 1 cf is 28.3 liters, we obtain that 1 mole of ideal gas
is 0.79 cf.

Thus, 4 moles = 4$\cdot$22.4 liters = 89.6 liters = 3.16 cf of
plasma require 1634.4 kcal energy input due to the above result.
So that we obtain the following estimation of the energy needed to
convert 1 mole of H$_2$O and 1 mole of C (graphite) to the plasma,
\be
  1634.4/4 = 408.5 \kcal
\ee
per one mole of the 100\% ionized 2H,O,C plasma; or
\be
 1634.4/3.16 = 517 \kcal
\ee
per one cubic foot of the 100\% ionized 2H,O,C plasma; or, using
the relation 1 kWh = 860 kcal,
\be
 517/860 = 0.6 \textrm{ kWh}
\ee
per one cubic foot of the 100\% ionized 2H,O,C plasma, or, using
the relation 1 kWh = 3413 BTU,
\be
 3413\cdot0.6 = 2052\textrm{ BTU}
\ee
per one cubic foot of the 100\% ionized 2H,O,C plasma.

The following remarks are in order. In our study, (i) we do not
take into account energies associated to cathode, anode, and in
the form of a radiation (DC electric low-voltage high-current
discharge in water vapor); (ii) we do not consider fluidodynamics
and thermodynamics issues associated to the flow and bubbles in
the PAF reactor; (iii) we do not consider mechanism of the
reactions and chemical kinetics issue; (iv) we do not consider
magnetochemistry (influence of strong external magnetic field on
the species and chemical reactions) of the PAF reactor; and (v) we
do not consider the creation of clusters containing molecules and
atoms.

\subsection{No recombination of water}\label{sec6.2}

We assume 1 mole of liquid water and 1 mole of solid graphite, at
$T=300$~K.

Reactions (\ref{2.1}) and (\ref{2.2}) below are due to the
formation of the 2H,O,C plasma, with $T=3300$ K, from the above
water and graphite. Reaction (\ref{2.3}) is ion recombination of
2H; reaction (\ref{2.4}) is formation of H$_2$ gas; reaction
(\ref{2.5}) is ion recombinations of C and O; and reaction
(\ref{2.6}) is formation of CO gas; namely,
\be
\label{2.1}
 \H_2\O(\textrm{liquid, 300 K}) \to
2\H(\textrm{plasma, 3300 K}) + O(\textrm{plasma, 3300 K}) - 1197
\kcal,
\ee
 where
\be
 Q_{\mathrm{water}}
= 1197 \kcal.
\ee
\be
\label{2.2}
 \C(\textrm{solid, 300K}) \to \C(\textrm{plasma,
3300K}) - 437.4 \kcal,
\ee
 where
\be
 Q_{\mathrm{carbon}}
= 437.4 \kcal.
\ee
\be
\label{2.3}
 2\H(\textrm{plasma, 3300K})  \to 2\H(\textrm{gas,
300K})  + 625 \kcal,
\ee
where 625 kcal= 2N13.6 eV = 163.2$\cdot10^{23}$ eV is ion
recombination heat of 2 moles of H (2H$^{+}  + 2e \to 2$H);
\be
\label{2.4}
 2\H(\textrm{gas, 3300K})  \to \H_2(\textrm{gas,
300K})  + 104.2 + 18 \kcal.
\ee
Here, 104.2 kcal are released due to recombination heat of H$_2$
molecule, H + H $\to$ H$_2$, and 18 kcal are due to cooling down
of a diatomic gas from $T=3300 K$ to $T=300 K$. Heat capacity of a
diatomic gas is about 6 to 7 cal/(mole K), at high and low
temperatures.
\be
\label{2.5}
 \C(\textrm{plasma, 3300K}) + \O(\textrm{plasma,
3300K})
 \to \C(\textrm{gas, 3300K}) +
\ee
$$
+ \O(\textrm{gas, 3300K}) + 259+313 \kcal;
$$
where 259 kcal= N11.26 eV = 67.6$\cdot$10$^{23}$ eV is ion
recombination heat of 1 mole of C (C$^{+} + e  \to $C), and 313
kcal = N13.6 eV = 81.6$\cdot$10$^{23}$ eV is ion recombination
heat of 1 mole of O (O$^{+} + e \to $O);
\be
\label{2.6}
 \C(\textrm{gas, 3300K}) + \O(\textrm{gas, 3300K})
\to \C\O(\textrm{gas, 300K}) + 255.8 + 18 \kcal,
\ee
where 255.8 is energy released during formation of carbon monoxide
CO, and 18 is energy released due to the cooling down of CO from
$T=3300 K$ to $T=300 K$.

In conclusion,

(i) The creation of 4 moles of fully ionized 2H,O,C plasma
(T=3300K) requires
\be
 Q_{\mathrm{water}}+Q_{\mathrm{carbon}} =
1197+437=1634 \kcal;
\ee
Thus, the energy consumption for the plasma is (1/4)1634 kcal/mol,
i.e.
\be
 408.5\textrm{ kcal/mol} = 0.475\textrm{ kWh/mol}
 = 1621\textrm{BTU/mol}  = 515.8\textrm{ kcal/cf} =
\ee
$$
= 0.600\textrm{ kWh/cf} = 2047\textrm{ BTU/cf of the
plasma};
$$

(ii) The formation of 1 mole of H$_2$(gas, 300K) releases
625+104.2 + 18 = 747 kcal;

(iii) The formation of 1 mole of CO(gas, 300K) releases
259+313+255.8 + 18 = 846 kcal;

(iv) In total, 747+846= 1593 kcal is released as a heat. Thus, the
heat released is (1/4)1593 =398 kcal/mol=1994 BTU/cf of the
plasma.

(v)  In total, 2 moles of the CO+H$_2$ (1:1 ratio) gas
have been
produced from 4 moles of the plasma (more precisely,
from 4 moles of
the positive ions and 4 moles of electrons);

(vi) As the net result, from (i) and (iv) we obtain 1593-1634 = -
41 kcal per two moles of CO+H$_2$ gas, i.e. the considered
reaction,
\be
\C\solid + \H_2\O\liquid \to \C\plasma+ 2\H\plasma +
\O\plasma \to
\ee
$$
\to \C\O\gas + \H_2\gas,
$$
is endothermic. Within the adopted accuracy, this value confirms
the value 43.9 kcal of Eq.~(\ref{water}) obtained without
consideration of the intermediate plasma state.

(vii)  Since the number of moles of the gas produced is two times
less than the number of moles of the plasma we have, in addition
to the above results (i) and (iv), the following alternative
result. The energy input is 2$\cdot$2047 = 4094 BTU/cf of the
CO+H$_2$ gas; and the heat produced by the exothermic reactions is
2$\cdot$1994 BTU/cf = 3988 BTU/cf of the CO+H$_2$ gas. The total
balance is $- 4094 + 3988 = -108$ BTU/cf.

\subsection{50\% recombination of water}\label{sec6.3}

The reaction
\be
\C + \H_2\O \to \C(\textrm{plasma})+
2\H(\textrm{plasma}) +
\O(\textrm{plasma}) \to \C\O + \H_2
\ee
represents an ideal situation because in reality some atoms may
recombine back into the water. Therefore, we should consider the
more general case,
\be
\C + x\H_2\O \to \C(\textrm{plasma})+ 2\H(\textrm{plasma}) +
\O(\textrm{plasma}) \to x_1\H_2\O + x_2\C\O + x_3\H_2.
\ee
For $x_1 \not= 0$, we have lower efficiency of the process since
the target products are CO and H$_2$.

Below, we consider the sequence starting with 1 mole of water and
1/2 mole of graphite; reactions (\ref{3.1}) and (\ref{3.2}) below
are formation of the 2H, O, (1/2)C plasma; reaction (\ref{3.3}) is
ion recombination of H and (1/2)O, and recombination of liquid
(1/2)H$_2$O (50\% recombination); reaction (\ref{3.4}) is ion
recombination of the remaining H; reaction (\ref{3.5}) is
formation of (1/2)H$_2$; reaction (\ref{3.6}) is ion recombination
of (1/2)C and remaining (1/2)O; and reaction (\ref{3.7}) is
formation of CO; namely,
\be
\label{3.1}
 \H_2\O(\textrm{liquid, 300K}) \to 2\H(\textrm{plasma,
3300K}) + \O(\textrm{plasma, 3300K}) - 1197 \kcal,
\ee
      i.e., the same as the above reaction
(\ref{2.1});
\be
\label{3.2}
 \frac{1}{2}\C(\textrm{solid, 300K}) \to
\frac{1}{2}\C(\textrm{plasma, 3300K}) - (1/2)437 \kcal,
\ee
      i.e., 218.5 (one-half of the above reaction
(\ref{2.2}));
\be
\label{3.3}
 2\H(\textrm{plasma, 3300K}) + \O(\textrm{plasma,
3300K}) \to
\frac{1}{2}\H_2\O(liquid, 300K) +
\ee
$$
+ \H(\textrm{plasma, 3300K}) + \frac{1}{2}\O(\textrm{plasma,
3300K}) + \frac{1}{2}1197 \kcal,
$$
      i.e., 598.5 kcal release;
\be
\label{3.4}
 \H(\textrm{plasma, 3300K})  \to \H(\textrm{gas,
300K})  + \frac{1}{2}625 \kcal,
\ee
      i.e., 312.5 kcal release;
\be
\label{3.5}
 \H(\textrm{gas, 3300K})  \to
\frac{1}{2}\H_2(\textrm{gas,
300K})
 + \frac{1}{2}104.2 \kcal + \frac{1}{2}18 \kcal,
\ee
      i.e., 52.1+9= 61.1 kcal release;
\be
\label{3.6}
 \frac{1}{2}\C(\textrm{plasma, 3300K}) +
(1/2)\O(\textrm{plasma, 3300K}) \to
 \frac{1}{2}\C(\textrm{gas, 3300K}) +
\ee
$$
+ \frac{1}{2}\O(\textrm{gas, 3300K}) + \frac{1}{2}(259+313) \kcal,
$$
      i.e., 129.5+156.5=286 kcal release.
\be
\label{3.7}
 \frac{1}{2}\C(\textrm{gas}, 3300K) +
\frac{1}{2}\O(\textrm{gas}, 3300K) \to
\frac{1}{2}\C\O(\textrm{gas}, 300K) + \frac{1}{2}255.8 +
\frac{1}{2}18 \kcal,
\ee
      i.e., 128+9=137 kcal release;

In conclusion,

(i) The formation of 3.5 moles of 2H, O, $\frac{1}{2}$C plasma
with T=3300K from 1 mole of liquid water and 1/2 mole of solid
graphite (T=300K) requires 1197+218.5= 1415.5 kcal. Thus, the
energy consumption for the plasma is (1/3.5)1415.5 kcal/mol, i.e.
404.4 kcal/mol = 0.47 kWh/mol = 1605 BTU/mol = 510.6 kcal/cf =
0.594 kWh/cf = 2026 BTU/cf of the plasma;

(ii) 50\% recombination of water  (1/2 moles of water) releases
598.5 kcal;

(iii) The formation of 1/2 moles of H$_2$ releases
$\frac{1}{2}(625+104.2+18) = 312.5+52.1+9=373.6$ kcal;

(iv) The formation of 1/2 moles of CO releases $\frac{1}{2}(259 +
313 +255.8 + 18) = 129.5+156.5+128+9=423$ kcal;

(v) In total, 1 mole of CO+H$_2$ (1:1) gas and 1/2 mole of water
has been produced from 3.5 moles of the 2H, O, $\frac{1}{2}$C
plasma (more precisely, from 3.5 moles of the positive ions and
3.5 moles of electrons);

(vi) In total, 598.5+373.6+423 = 1395.1 kcal released as a heat.
Thus, the heat released is $\frac{1}{3.5}$1395.1 =399 kcal/mol =
1997 BTU/cf of the plasma.

(vii) Since the number of moles of the gas produced is 3.5 times
less than number of moles of the plasma we alternatively have, in
addition to the above results (i) and (vi), that the energy input
is 3.5$\times$2026 = 7091 BTU/cf  of the CO+H$_2$ gas; and the
heat produced  by the exothermic reactions is 3.5$\times$1997
BTU/cf = 6990 BTU/cf of the CO+H$_2$ gas.

\section{Heat production}

\subsection{No heat production}

No heat production is here understood in the sense that all the
heat produced by the exothermic reactions is used back in the
endothermic reactions, and thus helps the formation of the plasma
and CO+H$_2$ gas. The energy balance could be calculated as
follows:

\be
\textit{Energy required to form the plasma + Energy released as a
heat.}
\ee
Since the energy input is negative while heat produced is positive
we obtain from the above result (vi) of Sec. \ref{sec6.2} the
following energy consumption:
\be
\label{4.1a}
 -4094+3988 = -106 \textrm{ BTU/cf of the gas};
\ee
and from the above result (vii) of Sec. \ref{sec6.3}
(the case of 50\%
recombination of water):
\be
\label{4.1b}
 -7091+6990 = -101 \textrm{ BTU/cf of the gas}.
\ee

Here, minus sign means that energy is required. Hence, about 100
BTU is required to produce 1 cf of the gas, under the assumption
that the reactor has ideal 100\% efficiency (does not produce any
heat but only the gas).

\subsection{The heat production coefficient}

The heat produced by the exothermic reactions (see (vi) of Sec.
\ref{sec6.2} and (vii) of Sec. \ref{sec6.3}) is distributed via
two main channels: first, it contributes to the endothermic
reactions and, second,  it is dissipated into the environment
(heat production). If some part $k$,
\be
\label{4.1}
 0 < k < 1,
\ee
of the heat produced by the exothermic reactions is removed due to
dissipation (convection, radiation, heat-mass transfer, etc.)  to
the environment from the region where the endothermic reactions
occur, i.e., the "heat production coefficient" is given by
\be
\label{4.2}
            k =\frac{\textrm{Heat transferred to
environment}}{\textrm{Heat produced by exothermic
reactions}}.
\ee
Therefore the remaining part, $(1-k)$, of the heat,
\be
\label{4.3}
 1- k =\frac{\textrm{Heat transferred to endothermic
reactions}}{\textrm{Heat produced by exothermic
reactions}}
\ee
is used in the endothermic reactions. The latter part of heat
cannot be measured directly since it is absorbed by the
endothermic reactions thus helping the formation of the plasma and
CO+H$_2$ gas. Therefore, we could modify the above energy
consumptions (\ref{4.1a}) and (\ref{4.1b}) as follows:
\be
\label{4.4a}
 -4094+(1-k)3988 \textrm{ BTU/cf}
\ee
 and
\be
\label{4.4b}
 -7091+(1-k)6990 \textrm{ BTU/cf}
\ee
of the gas, with the associated heat produced being
\be
\label{4.5a} k3988 \textrm{ BTU/cf}
\ee
and
\be
\label{4.5b}
 k6990 \textrm{ BTU/cf}
\ee
of the gas, respectively. The heat productions (\ref{4.5a}) and
(\ref{4.5b}) are those corresponding to the measurable heat
produced since these heats are absorbed by the environment
(surrounding liquid, metal parts of the reactor, etc.).


\subsection{Example 1: 70\% heat production}

For k=0.7 (70\% of the total heat is dissipated/utilized and 30\%
is used in the endothermic reactions), we get energy consumptions
\be
\label{4.6a} -4094+0.3\cdot3988 = -2898 \textrm{
BTU/cf}
\ee
and
\be
\label{4.6b}
 -7091+0.3\cdot6990 = -4994 \textrm{ BTU/cf}
\ee
of the gas, with the associated heat production (i.e. measurable
heat produced) being
\be
\label{4.7a} 0.7\cdot3988 = 2792 \textrm{ BTU/cf}
\ee
of the gas and
\be
\label{4.7b} 0.7\cdot6990 = 4893 \textrm{ BTU/cf}
\ee
of the gas, respectively. These heat productions correspond to
measurable heats produced.

\subsection{Example 2: 100\% heat production}

For k=1 (100\% of the heat produced by exothermic reactions is
dissipated/utilized), we evidently have the maximal value for the
energy consumptions:
\be
\label{4.8a}
 -4094 \textrm{ BTU/cf}
\ee
of the gas; and
\be
\label{4.8b}
 -7091 \textrm{ BTU/cf}
\ee
of the gas, and the associated maximal values of the heat
productions:
\be
\label{4.9a}
 3988 \textrm{ BTU/cf}
\ee
of the gas; and
\be
\label{4.9b}
 6990 \textrm{ BTU/cf}
\ee
of the gas, respectively. These heat productions correspond to
measurable heats produced.

\section{Total heat produced}

We now add the combustion heat of the produced CO+H$_2$ (1:1) gas
(the theoretical value is 315 BTU/cf) to the measurable heat
produced by the reactor, in order to estimate the total heat
produced. By adding 315 BTU/cf to the heats (\ref{4.4a}) and
(\ref{4.4b}) we obtain the total measurable heat produced
\be
\label{5.1a}
 k3988 + 315 \textrm{ BTU/cf of the CO+H$_2$ gas;}
\ee
and
\be
\label{5.1b}
 k6990 + 315 \textrm{ BTU/cf of the CO+H$_2$ gas,}
\ee
for the cases of 0\% and 50\% recombination of water,
respectively. Here, the coefficient $k$ ($0<k<1$) is defined by
(\ref{4.2}) and can be given in some approximate value by studying
thermodynamics of a specific reactor. This coefficient accounts
for all heat losses, including that at (tungsten) anode.

By assuming that the total heat produced is approximately equal to
the energy input (see the energy inputs in (vi) of Sec.
\ref{sec6.2} and (vii) of Sec. \ref{sec6.3}), that is, by assuming
the efficiency 1, we have
\be
\label{5.2a}
 \frac{k3988 + 315\textrm{ BTU/cf of the CO+H$_2$
gas}}{4094\textrm{ BTU/cf of the CO+H$_2$ gas}} = 1,
\ee
and
\be
\label{5.2b}
\frac{k6990 + 315\textrm{ BTU/cf of the
}CO+H_2\textrm{
gas}}{7091 \textrm{ BTU/cf of the CO+H$_2$ gas}} = 1.
\ee
Therefore, we obtain
\be
\label{5.3a}
 k=0.948
\ee
and
\be
\label{5.3b}
 k= 0.969,
\ee
respectively. The measurable total heat produced is
\be
\label{5.4a}
 4094\textrm{ BTU/cf of the CO+H$_2$ gas};
\ee
and
\be
\label{5.4b}
 7091\textrm{ BTU/cf of the CO+H$_2$ gas},
\ee
which is valid under conditions (\ref{5.2a}) and (\ref{5.2b}),
i.e., that the total measurable heat produced is equal to the
energy input.

The above estimations (\ref{5.3a}) and (\ref{5.3b}) mean that
about 95\% of the (electric) energy input is dissipated into the
environment and the remaining 5\% contributes to the endothermic
chemical reactions.

The following remark is in order. We can account for additional
heat production which could not be accounted by the coefficient
$k$ by adding some heat $Q'$, so that (\ref{5.4a}) and
(\ref{5.4b}) become
\be
\label{5.5a}
 \frac{k3988 + 315 + Q'\textrm{ BTU/cf of the CO+H$_2$
gas}}{4094\textrm{ BTU/cf of the CO+H$_2$ gas}}= A,
\ee
and
\be
\label{5.5b}
 \frac{k6990 + 315 + Q'\textrm{ BTU/cf of the CO+H$_2$
gas}} {7091 \textrm{ BTU/cf of the CO+H$_2$ gas}}= A,
\ee
where $A$ can be taken approximately one, or some other value.
However, one can incorporate $Q'$ into $k$ by simple redifinition.
For instance, $k3988 + 315 + Q' \to k3988 + 315 + k'3988 \to
(k+k')3988 + 315 \to k3988 + 315$, and we arrive again to the
estimation (\ref{5.3a}), for $A=1$. However, here $k$ acquires
some other meaning which is different than that in Eq.
(\ref{4.2}).

\section{Conclusions}

 In this paper we have studied the upper limit of the "commercial
efficiency" \cite{1}, simply referred i  the text as "efficiency"
of Santilli's hadronic reactors of molecular type, also called
PlasmaAArcFlow reactors \cite{1}.

 For the case of pure water we have obtained the upper limit 3.11,
while for the mixture of ethyleneglicole and water the efficiency
is given by Eq. (\ref{rless}) (see Figs.~1 and 2), and rises from
3.11 to about 7.5, with increase of the relative consumption of
ethyleneglicole with respect to that of carbon rods. These results
are based on the assumptions that all thermochemical processes are
conventional and have been obtained at at $T = 25^o$C and $p=1$
atm.

 Similar calculations can be made for different water-based liquid
wastes, provided that their main chemical
composition is known.

 We have accounted for the plasma production, and obtained a more
detailed view on the steps of the entire process. This has allowed
us to account for the water recombination and the heat production
coefficient.

The comparison of the theoretical results with measurements leads
to a sharp discrepancy between theoetical predictions and
experimental data. In fact, the fitting of the measured commercial
over-unity leads to a prediction which is about three times
smaller than the measured heat. On the other hand, the fitting of
the measured heat production leads to the prediction that the
measured commercial over-unity is smaller than that predicted.

The implications of the above calculations are the following. The
fact that the heat produced in the PlasmaArcFlow reactors is less
than 1/3 the theoretical prediction constitutes clear evidence
that magnegas is not composed of conventional H$_2$ and CO
molecules. Alternatively, the indicated evidence prohibits the
complete formation of H$_2$ and CO. It then follows that magnegas
is indeed composed of clusters, called Santilli magnecules
\cite{1}, which  are composed of clusters of individual H, C and O
atoms, dimers OH, CH or C-O in single bond, and ordinary molecules
$H_2$ and CO under a new attraction between opposite polarities
created by magnetic and electric polarizations of the orbitals of
individual atoms into toroids.

On the other side, the fact that the heat content of magnegas is
about 3 times that predicted is additional strong evidence that
magnegas, again, contains non-molecular bonds. In fact, said H, C
and O atoms may bond into H$_2$ and CO at the time of combustion,
thus releasing extra energy. As such the above  two large
deviations complement each other rather nicely.

It is evident that no additional quantitative study of the
commercial efficiency of  PlasmaArcFlow reactors can be done
without a more accurate knowledge of  the new chemical species of
magnecules, as well as a reinterpretation of thermochemical
processes via the covering hadronic chemistry.


\appendix

\section*{Appendix}

\subsection*{Hydrogen}

 Atomic weight: 1 gram/mol;\\
 Ionization potential: 13.6 eV;\\
 Melting point: 13.8 K;\\
 Boiling point: 20.3 K;\\
 Specific heat capacity (300K): 14.304 J/(gram K) = 3.4~cal/(gram
K) = 3.4~cal/(mole K);\\
 Density: 0.09 gram/liter (gas).

To heat 1 mole (i.e. 1 gram) of hydrogen from $T=300$K to
$T=3300$K it is required 3.4 cal/(mole K)$\cdot$(3300 - 300) K =
10.2 kcal.

\subsection*{Oxygen}

\noindent Atomic weight: 16 gram/mol;
 First Ionization potential: 13.6 eV;\\
 Melting point: 54.8 K;\\
 Boiling point: 90.2 K;\\
 Specific heat capacity (300K): 0.92 J/(gram K) = 0.22 cal/(gram K) =
 3.5~cal/(mole K).

To heat 1 mole (i.e. 16 grams) of oxygen from $T=300$K to
$T=3300$K it is required 3.5 cal/(mole K )$\times$(3300 - 300) K =
10.5 kcal.

\subsection*{Carbon}

\noindent
 Atomic weight: 12 gram/mol;\\
 First ionization potential: 11.26 eV;\\
 Melting point: 3825 K;\\
 Boiling point:  5100 K;\\
 Specific heat capacity (300K): 0.709 J/(gram K) = 0.17 cal/(gram
K) = 2~cal/(mole K);\\
 Heat of evaporation: 715 kJ/mol = 171.7 kcal/mol 1 J = 0.24 cal 1
J = 2.8$\times 10^{-7}$ kWh 1 J = $10^{7}$ erg.

To heat 1 mole (i.e., 12 grams) of carbon from $T=300$K to
$T=3300$K it is required 2~cal/(mole K )$\times$(3300 - 300) K =
6000 cal = 6 kcal, thus $E_1$ = 6 kcal.

\subsection*{Water (liquid)}

\noindent
 Molecular weight: 18 gram/mol;\\
 Specific heat capacity (300K): 4.18 J/(gram K) = 1 cal/(gram K) =
18~cal/(mole K).

 To heat 1 mole (i.e., 18 grams) of liquid water from $T=20$C to
$T=100$C it is required 18 cal/(mole K )$\times$(100 - 20) K =
1440 cal = 1.4 kcal, thus $Q_1$ = 1.4 kcal.

\subsection*{Water vapor (ideal gas)}

\noindent Molecular weight: 18 gram/mol;

For ideal gases the heat capacity is $Nk/2$ per each degree of
freedom of molecule. $Nk/2$ = 4.2 J/(mole K) = 1 cal/(mol K).

Water molecule has 3 translational and 3 rotational degrees of
freedom. Also, there are 3 vibrational degrees of freedom, at
sufficiently high temperatures ($T >$ 3000K). So, in total we have
(about) 9 degrees of freedom. Hence, the heat capacity of water
vapor at high temperatures is $9Nk/2$ = 9 cal/(mol K). The heat
required to heat up 1 mole of water vapor from T =100C = 400K to
$T=3300$K is thus 9 cal/(mole K)$\times$(3300 - 400) K = 26100 cal
= 26 kcal, therefore $Q_3$ = 26 kcal.

\subsection*{Conversion factors and constants}

\noindent
 1 kWh = 860 kcal = 3413 BTU;\\
 1 kcal = 3.97 BTU;\\
 1 eV =3.83 x 10$^{-23}$ kcal;\\
 1 cal = 4.18 J;\\
  1 mole = 22.4 liters = 0.792 cf (an
ideal gas, normal conditions);\\
 1 cf = 28.3 liters;\\
 1 cf = 1.263 mol (an ideal gas, normal conditions);\\
 $N$ = 6$\times$10$^{23}$ mol$^{-1}$ (Avogadro number);\\
 $Nk/2$ = 1 cal/(mol K);\\
 $R$ = 8.314 J/(mol K) = 1.986 cal/(mol K).

\subsection*{Specific heat capacities}

\noindent $p = 1$ atm, $T = 25$C.

\noindent
 H$_2$(gas): 29.83 J/(mol K) = 7 cal/(mol K);\\
 H$_2$O (liquid): 4.18 J/(gram K) = 1 cal/(gram K)
= 18 cal/(mol~K);\\
 C (graphite, solid): 0.71 J/(gram K) = 0.17
cal/(gram K) = 2~cal/(mol~K);\\
 O$_2$ (gas): 29.36 J/(gram K) = 7 cal/(gram K);\\
 H (gas): 14.3 J/(gram K) = 3.42 cal/(gram K);\\
 O (gas): 0.92 J/(gram K) = 0.22 cal/(gram K);\\
 Fe (solid): 0.45 J/(gram K) = 0.11 cal/(gram K) =
6 cal/(mol K).

\subsection*{Evaporation heats}

\noindent
  Water:   10.4 kcal/mol, $T=25$C;\\
  Graphite:   171.7 kcal/mol, $T=25$C.


\subsection*{First ionization potentials}

 H:   13.6 eV;
 O:   13.6 eV;
 C:   11.26 eV.

\subsection*{Test results for some model of PAF
reactor}

\noindent Measured energy consumption: 100 Wh/cf = 341.3 BTU/cf of
magnegas;\\
 Measured heat production: 665 BTU/cf of magnegas
produced;\\
 Measured combustion heat of magnegas: 650 BTU/cf;\\
Measured commercial over-unity: (665+650)/341.3 = 3.85;\\
Theoretical combustion heat of CO+H$_2$ (1:1) gas: 315 BTU/cf.


\newpage
\begin{thebibliography}{6}

\bibitem{1}
{\tt http://www.usmagnegas.com}

\bibitem{2}
R.M. Santilli, {\it Foundations of Hadronic Chemistry with
Aopplications to New Clean Energies and Fuels} (Kluwer Academic
Publisher, Boston-Dordrecht-London, 2001).

\bibitem{3}
R.M. Santilli and D.D. Shillady, {\it Ab Initio Hadronic
Chemistry}, Hadronic Press, Florida (2000).

\bibitem{4}
R.M. Santilli and D.D. Shillady, International Journal of Hydrogen
Energy {\bf 24}, 943 (1999), and {\bf 25}, 173 (2000).

\bibitem{5}
 (a) M.G.~Kucherenko and A.K.~Aringazin, Hadronic Journal {\bf 21},
895 (1998);
 (b) M.G.~Kucherenko and A.K.~Aringazin, Hadronic
Journal {\bf 23}, 1 (2000). e-print http://www.arXive.org: {\tt
physics/0001056};
 (c) A.K.~Aringazin, Hadronic Journal  {\bf 23},
57 (2000). e-print http://www.arXive.org: {\tt physics/0001057}.

\bibitem{6}
R.M. Santilli, "Alarming oxygen depletion caused by hydrogen
combustion and their resolution by magnegas\tm{}", Contributed
paper, International Hydrogen Energy Forum 2000, Munich, Germany,
September 11-15, 2000. e-print http://www.arXive.org: {\tt
physics/0009014}.
\end{thebibliography}
\end{document}